# Theoretical description of the first-order phase transition of aluminum from a superconducting to normal state by the current-density functional theory for superconductors

Katsuhiko Higuchi[1] and Masahiko Higuchi[2]

[1]Graduate School of Advanced Sciences of Matter, Hiroshima University, Higashi-Hiroshima 739-8527, Japan
[2]Department of Physics, Faculty of Science, Shinshu University, Matsumoto 390-8621, Japan

*Abstract*

We show that the current-density functional theory for superconductors (sc-CDFT) can describe the magnetic-field-induced first-order phase transition of aluminum from a superconducting state to a normal state. This is accomplished by introducing a model for the magnetic field dependence of the attractive interaction between superconducting electrons. This model states that the surface potential well produced by the penetrating magnetic field leads to the magnetic field dependence of the attractive interaction. Specifically, the electron density near the surface increases with the magnetic field due to the surface potential well, which causes the reduction in the attractive interaction due to the screening effect. We also develop the calculation scheme to solve the gap equation of the sc-CDFT with taking into account the magnetic field dependence of the attractive interaction. It is shown that calculation results of the magnetic field dependence of the superconducting gap are in good agreement with experimental results of the first-order phase transition.

# I. INTRODUCTION

The density functional theory for superconductors (sc-DFT) is widely used for predicting the thermal equilibrium properties of superconductors [1-31]. The sc-DFT is extended so that one can treat superconductors immersed in a magnetic field [32-39]. The current-density functional theory for superconductors immersed in a magnetic field was first developed by W. Kohn el al. [32], in which they pointed out the possibility to describe the Meissner effect quantitatively in conjunction with Maxwell's equations. We have recently developed the current-density functional theory for superconductors immersed in a magnetic field (sc-CDFT) [33,37-39] on the basis of the extended constrained-search theory [40-44]. In the sc-CDFT, the electron density, the transverse component of the paramagnetic current density, the spin current density, the superconducting order parameter (OPSS) and its complex conjugate are chosen as the basic variables [37-41]. Thermal equilibrium values of basic variables can be reproduced by means of solutions of a so-called Bogoliubov-de Gennes-Kohn-Sham (BdG-KS) equation. Using the reproduced basic variables, we can calculate the charge density and current density. By substituting the obtained charge and current densities into Maxwell's equations and solving them, the scalar and vector potentials inside the superconductor are obtained. Substituting the obtained scalar and vector potentials into the BdG-KS equation, we get the charge density and current density again. By continuing the above process until it becomes self-consistent, the BdG-KS equation and Maxwell's equations are solved simultaneously [32,33,37-39]. Thus, it is possible not only to get the charge distribution, the current density distribution, and the OPSS distribution in the superconductor, but also to determine the scalar and vector potentials. In this way, the Meissner effect can be described in the sc-CDFT [32,33,37-39].

In order to solve the BdG-KS equation of the sc-CDFT, we have proposed an approximation method [37,38] in which the solutions of the BdG-KS equation are supposed to have the same spatial dependence as the single-particle wave functions of the normal state [45]. By using this approximation, the problem of solving the BdG-KS equation is reduced to the problem of finding the eigenvalues and eigenvectors of a 2x2 matrix [37,38]. Furthermore, the approximate form of the exchange-correlation energy functional of the sc-CDFT is developed by applying the mean field approximation to the attractive interaction part of the exchange correlation energy functional [37-39]. By using the approximate form, the problem of finding the eigenvalues and eigenvectors of a 2x2 matrix is further reduced to the problem of solving the gap equation for superconductors immersed in a magnetic field [38,39]. As a result, the problem of solving the BdG-KS equation and Maxwell's equations simultaneously is reduced to the problem of solving the gap equation of the sc-CDFT and Maxwell's equations simultaneously [38,39].

Furthermore, in order to perform actual calculations on the basis of the sc-DFT, we introduce the assumption that the magnetic field distribution obtained in the London theory [46] corresponds to that obtained by solving the BdG-KS equation and Maxwell's equations simultaneously. Under this assumption, we developed a calculation scheme in which the superconducting gap and attractive interaction are treated as variables that are determined by solving the gap equation of the sc-CDFT in conjunction with the energy balance equation [39]. The energy balance equation states that the superconducting gap in a magnetic field corresponds to the energy gain of superconducting state electrons against normal state electrons and is equal to the energy gain at zero magnetic field minus the potential energy of the diamagnetic magnetization [38,39]. It is found that the attractive interaction decreases as the magnetic field increases, which leads to the decrease in the superconducting gap. It is known that the superconducting gap exhibits a first-order transition with respect to magnetic field and temperature [47,48]. However, the previous scheme [39] cannot describe the experimentally observed first-order phase transition in aluminum immersed in a magnetic field [47,48]. The reason for this discrepancy would be that the attractive interaction obtained by solving the gap equation of the sc-CDFT in conjunction with the energy balance equation may be unrealistic value [39].

In this paper, in order to describe the first-order transition observed in aluminum [47,48], we propose a model for the magnetic field dependence of the attractive interaction and also develop the calculation scheme in which the magnetic field dependence is taken into account. It is shown that the present calculation scheme that includes the mechanism of the reduction in the attractive interaction can successfully described the first-order transition observed in aluminum immersed in a magnetic field.

The rest of the paper is organized as follows. In Sec. II A, the outline of the sc-CDFT is presented. In Sec. II B, we explain the model for the magnetic field dependence of the attractive interaction. The calculation scheme is explained in Sec. II C. In Sec. III, we give the calculation results and discuss how the first-order phase transition can be described in the present scheme. Finally, some concluding remarks are given in Sec. IV.

## II. CALCULATION SCHEME
### A. sc-CDFT and its application to aluminum immersed in a magnetic field

In the sc-CDFT, the electron density, transverse component of the paramagnetic current density, and spin current density, OPSS and its complex conjugate are chosen as basic variables [38,39]. The basic variables for the equilibrium state are calculated by using the solution of the BdG-KS equation. The BdG-KS equation of the sc-CDFT is given by [38,39]

$$\begin{aligned}(h_s^r - \mu)u_k(r\zeta) + \int\{D_s(r\zeta,r'\zeta') - D_s(r'\zeta',r\zeta)\}v_k^*(r'\zeta')\mathrm{d}r'\mathrm{d}\zeta' &= E_k u_k(r\zeta)\\ -(h_s^r - \mu)v_k(r\zeta) + \int\{D_s(r\zeta,r'\zeta') - D_s(r'\zeta',r\zeta)\}u_k^*(r'\zeta')\mathrm{d}r'\mathrm{d}\zeta' &= E_k v_k(r\zeta),\end{aligned} \quad (1)$$

where $h_s^r$ denotes the single particle Hamiltonian that is given by

$$h_s^r = \frac{\{p + eA_s(r)\}^2}{2m} + v_s(r) + g\frac{\mu_B}{\hbar}\hat{s}_{op}^\zeta \cdot \nabla \times A_s(r). \quad (2)$$

In Eqs. (1) and (2), $v_s(r)$, $A_s(r)$, $D_s(r\zeta,r'\zeta')$ and $D_s^*(r\zeta,r'\zeta')$ are the effective potentials that are determined so that the solutions of the BdG-KS equation, $u_k(r\zeta)$ and $v_k(r\zeta)$, reproduce basic variables for the equilibrium state [38,39]. Specific expressions for effective potentials include the exchange-correlation energy functional of the sc-CDFT [38, 39]. This functional contains both the exchange-correlation effect of the electron–electron Coulomb interaction and attractive interaction between superconducting electrons. Concerning the exchange-correlation energy functional, we have developed the approximate form by using the mean field approximation [38, 39].

In this paper, the sc-CDFT is applied to an aluminum plate immersed in a magnetic field that is parallel to the $z$-axis and is denoted as $(0,0,B)$. Figure 1(a) shows the schematic diagram of the system. The thickness of the aluminum plate is $L_x$ and expands in $y$ and $z$ directions. The sizes of the system for $y$ and $z$ directions are denoted as $L_y$ and $L_z$, respectively. The periodic boundary condition is imposed on the single-particle wave functions of the normal state with the period of $L_y$ and $L_z$ in $y$ and $z$ directions, respectively. For simplicity, the homogeneous electron gas with $r_s$ =2.07 is considered, where the value of 2.07 for $r_s$ corresponds to the density of conduction electrons in aluminum [49].

Similarly to the previous works [38,39], we adopt the approximation method proposed by de Gennes [45] in solving the BdG-KS equation. Namely, $u_k(r\zeta)$ and $v_k(r\zeta)$ are given by the multiplication of the single-particle wave functions of the normal state, $w_{k\sigma}(r)\chi_\sigma(\zeta)$, and constant coefficients, i.e., we have

$$\begin{aligned}u_k(r\zeta) &= \bar{u}_k w_{k\downarrow}(r)\chi_\downarrow(\zeta)\\ v_k(r\zeta) &= \bar{v}_k w_{k\uparrow}(r)\chi_\uparrow(\zeta),\end{aligned} \quad (3)$$

where $\bar{u}_k$ and $\bar{v}_k$ denote constant coefficients, and where the normal state $w_{k\sigma}(r)\chi_\sigma(\zeta)$ obeys the following equation:

$$(h_s^r - \mu) w_{k\sigma}(\mathbf{r}) \chi_\sigma(\zeta) = \xi_{k\sigma} w_{k\sigma}(\mathbf{r}) \chi_\sigma(\zeta). \tag{4}$$

As in the previous works [38,39], we suppose that $A_s(\mathbf{r})$, which should be determined by solving the BdG equation in conjunction with Maxwell's equations [32,33,37-39], is the same as the profile given by the London theory [46]. Namely, $A_s(\mathbf{r})$ is supposed to be given by

$$A_s(\mathbf{r}) = (0, \lambda \bar{B} \sinh(x/\lambda), 0) \tag{5}$$

with $\bar{B} = B/\cosh(L_x/2\lambda)$ [38,39]. Using Eq. (5), and using $v_s(\mathbf{r}) \approx 0$, $h_s^r$ can be rewritten by

$$h_s^r = \frac{p^2}{2m} + \frac{e\bar{B}\lambda}{m}\sinh\left(\frac{x}{\lambda}\right) p_y + \frac{e^2\bar{B}^2\lambda^2}{2m}\sinh^2\left(\frac{x}{\lambda}\right) + \frac{e\hbar\bar{B}}{2m}\bar{\sigma}\cosh\left(\frac{x}{\lambda}\right), \tag{6}$$

where $\bar{\sigma}$ is equal to 1 or -1 for a spin-up or spin-down state, respectively. Since $[h_s^r, p_y] = [h_s^r, p_z] = [p_y, p_z] = 0$, the eigenfunction of $h_s^r$ is given by [39]:

$$w_{k\sigma}(\mathbf{r})\chi_\sigma(\zeta) = \varphi_{k\sigma}(x) e^{i(k_y y + k_z z)} \chi_\sigma(\zeta), \tag{7}$$

Substituting Eq. (7) into Eq. (4), we get the following equation for $\varphi_{k\sigma}(x)$:

$$\left\{ -\frac{\hbar^2}{2m}\frac{d^2}{dx^2} + V_{\sigma,k_y}(x) - \mu \right\} \varphi_{k\sigma}(x) = \left\{ \xi_{k\sigma} - \frac{\hbar^2}{2m}(k_y^2 + k_z^2) \right\} \varphi_{k\sigma}(x) \tag{8}$$

with

$$V_{\sigma,k_y}(x) = \frac{e\hbar \bar{B} k_y \lambda}{m}\sinh(x/\lambda) + \frac{e\hbar\bar{B}}{2m}\bar{\sigma}\cosh(x/\lambda) + \frac{e^2\bar{B}^2\lambda^2}{2m}\sinh^2(x/\lambda). \tag{9}$$

Figure 1(b) shows the surface potential well $V_{\sigma,k_y}(x)$ that is formed due to the effective vector potential $A_s(\mathbf{r})$. The leading term of Eq. (9) is the first one that depends on $k_y$, $B$ and $\lambda$ [39]. Equation (9) indicates that electrons with positive $k_y$ tend to be accumulated on the left-hand side of the plate and vice versa for electrons with negative $k_y$ (see, Fig. 1(b)), which would cause the Meissner effect [39]. It is also found from Fig. 1(b) and Eq. (9) that the depth of the surface potential well increases with increasing $B$. In other words, the electron density near the surface increases with increasing $B$. This means that the attractive interaction is reduced near the surface due to the screening effect. In the next subsection, we will discuss the magnetic field dependence of the attractive interaction caused by the surface potential well.

As mentioned in Sec. I, if we adopt the approximation method proposed by de Gennes [45], then we can rewrite the problem of solving the BdG equation by that of solving the gap equation of the sc-CDFT. The gap equation is given by [37-39]

$$1 = V_0(B) D_B(\varepsilon_F) \int_0^{\hbar\omega_D} \frac{\tanh\left\{\frac{\beta}{2}\left(\sqrt{\xi^2 + |\Gamma_0(B,T)|^2} - \frac{\Delta\varepsilon(B)}{2}\right)\right\}}{\sqrt{\xi^2 + |\Gamma_0(B,T)|^2}} d\xi, \quad (10)$$

where $\Gamma_0(B,T)$, $V_0(B)$, $D_B(\varepsilon_F)$, $\Delta\varepsilon(B)$ and $\omega_D$ denote the superconducting gap, the attractive interaction between superconducting electrons, the density of states for normal states at the Fermi energy, the average energy splitting between up and down spin states and Debye frequency, respectively.

We have developed a practical scheme, in which the gap equation is solved in conjunction with the energy balance equation [38,39]. The energy balance equation is given by

$$2n_{\max}^{(2)}(B,T)\Gamma_0(B,T) = 2n_{\max}^{(2)}(0,T)\Gamma_0(0,T) - \frac{B^2}{2\mu_0}\left\{1 - \frac{2\lambda}{L_x}\tanh\left(\frac{L_x}{2\lambda}\right)\right\}\Omega, \quad (11)$$

where $2n_{\max}^{(2)}(B,T)$ and $\Omega$ denote the total number of superconducting electrons and the volume of the system ($\Omega = L_x L_y L_z$), respectively. The value of $2n_{\max}^{(2)}(B,T)$ can be calculated from the OPSS and is given by [37-39]

$$2n_{\max}^{(2)}(B,T) \approx \frac{D_B(\varepsilon_F)}{4} \int_0^{\hbar\omega_D} \frac{|\Gamma_0(B,T)|^2}{\xi^2 + |\Gamma_0(B,T)|^2} \tanh^2\left\{\frac{\beta}{2}\left(\sqrt{\xi^2 + |\Gamma_0(B,T)|^2} - \frac{\Delta\varepsilon(B)}{2}\right)\right\} d\xi. \quad (12)$$

Equation (11) represents the relation between energy gains of the superconducting state in the zero and nonzero magnetic field cases [38,39]. Equations (10) and (11) are utilized in actual calculations, the specific procedure of which will be explained in Sec. II C.

**B. Model for the attractive interaction in a magnetic field**

In this subsection, we propose a model for the magnetic field dependence of the attractive interaction $V_0(B)$. Supposing that the attractive interaction is approximately independent of the wave number due to the strong screening effect, the magnitude of the attractive interaction at zero magnetic field, $V_0(0)$, is inversely proportional to the fourth power of the screening wave number $\alpha$ and to the volume $\Omega$. Since the screening wave number $\alpha$ is given as the ratio of the plasma frequency and fermi velocity [50], $V_0(0)$ depends on the density of conduction electrons. Suppose that in the absence of a magnetic field, the conduction electrons are uniformly distributed, and their density is given as $n_{cond}$. In this case, $V_0(0)$ is inversely proportional to $n_{cond}^{3/2}$ and to the volume $\Omega$. If the proportionality constant denoted by $U$, then $V_0(0)$ can be rewritten as

$$V_0(0) = \frac{U}{\Omega n_{cond}^{3/2}}, \tag{13}$$

where $U$ is supposed to be independent of $B$.

In the case of $B \neq 0$, the bounded states are formed near the surface due to the surface potential well. The number of bounded states is denoted as $2N_{bound}(B)$ on both sides of surface. In order to describe the magnetic field dependence of the attractive interaction, a simple model is introduced. Figure 2 shows the schematic diagram of the model. The width of the surface potential well is supposed to be $\lambda$. In addition, bounded electrons are supposed to be uniformly distributed in the surface potential well. If electrons other than bounded ones are distributed uniformly in the system, then the total number of electrons existing in the surface potential well is given by $2N_{bound}(B)\left(1 - \frac{2\lambda}{L_x}\right) + 2\lambda L_y L_z n_{cond}$. Therefore, the electron density in the surface potential well is approximately given by

$$n_{well}(B) = \left\{ n_{cond} - \frac{2N_{bound}(B)}{L_x L_y L_z} \right\} + \frac{N_{bound}(B)}{\lambda L_y L_z}. \tag{14}$$

This density $n_{well}(B)$ contributes to the screening of the attractive interaction in the region of the surface potential well. By using Eq. (13), the attractive interaction for superconducting electrons in the surface potential well region $V_0(B)_{well}$ is given by

$$V_0(B)_{well} = U \frac{n_{well}(B)^{-2/3}}{\lambda L_y L_z}$$
$$= V_0(0) \frac{L_x}{\lambda} \left\{ \frac{n_{cond}}{n_{well}(B)} \right\}^{2/3}. \tag{15}$$

On the other hand, the number of electrons spread over the entire region is given by $n_{cond} L_x L_y L_z - 2N_{bound}(B)$. Therefore, the electron density in the bulk region is given by

$$n_{bulk}(B) = n_{cond} - \frac{2N_{bound}(B)}{L_x L_y L_z}. \tag{16}$$

Thus, for superconducting electrons in the bulk region, the attractive interaction is given by

$$V_0(B)_{bulk} = U \frac{n_{bulk}(B)^{-2/3}}{(L_x - 2\lambda) L_y L_z}$$
$$= V_0(0) \frac{L_x}{L_x - 2\lambda} \left\{ \frac{n_{cond}}{n_{bulk}(B)} \right\}^{2/3}, \tag{17}$$

where Eq. (13) is used.

As mentioned above, the attraction interaction is different between the superconducting electrons near the surface and those in the bulk region. This is because surface potential wells produce different electron densities near the surface and in the bulk, which in turn produces different screening effects. If the number of superconducting electrons in the surface potential well region is denoted by $N_s(B,T)$ on each side, then the attractive interaction for $2N_s(B,T)$ superconducting electrons is given by Eq. (15), while for $2n_{max}^{(2)}(B,T) - 2N_s(B,T)$ superconducting electrons the attractive interaction is given by Eq. (17). Since the attractive interaction depends on the position, we introduce the following average attractive interaction as a model for the magnetic field dependence of the attractive interaction:

$$V_0(B) = V_0(B)_{well} \frac{2N_s(B,T)}{2n_{max}^{(2)}(B,T)} + V_0(B)_{bulk} \left( \frac{2n_{max}^{(2)}(B,T) - 2N_s(B,T)}{2n_{max}^{(2)}(B,T)} \right)$$
$$= V_0(0) \left[ \left\{ \frac{n_{cond}}{n_{well}(B)} \right\}^{2/3} \frac{L_x}{\lambda} \frac{N_s(B,T)}{n_{max}^{(2)}(B,T)} + \left\{ \frac{n_{cond}}{n_{bulk}(B)} \right\}^{2/3} \frac{L_x}{L_x - 2\lambda} \left\{ 1 - \frac{N_s(B,T)}{n_{max}^{(2)}(B,T)} \right\} \right]. \tag{18}$$

This model is employed in solving the gap equation (10).

In order to solve the gap equation (10) with the attractive interaction of Eq. (18), we need the value of $N_s(B,T)$. Supposing that superconducting electrons near the surface are distributed uniformly in the width of $\lambda$, then the density of the superconducting electron $n_s(B,T)$ is given by

$$n_s(B,T) = \frac{N_s(B,T)}{\lambda L_y L_z}. \tag{19}$$

If $\lambda$ is given as the London penetration depth, then we have

$$N_s(B,T) = \frac{mL_y L_z}{\mu_0 e^2 \lambda}. \tag{20}$$

In actual calculations, we utilize Eq. (20) to get the value of $N_s(B,T)$. Specific calculation procedure will be discussed in the next subsection.

**C. Calculation procedure**

The calculation procedure is shown in Fig. 3. First, we set values of $B$, $T$ and $\lambda$ (steps (i) and (ii)). In step (iii), $N_s(B,T)$ is calculated by using Eq. (20). Then, in step (iv), we give a trial vale of $2n_{\max}^{(2)}(B,T)$ that appear in Eq. (18). In step (v), we solve Eq. (8) to get $D_B(\varepsilon_F)$, $\Delta\varepsilon(B)$ and $N_{bound}(B)$. In solving Eq. (8), we utilize the perturbation theory, where $V_{\sigma,k_y}(x)$ is treated as a perturbation potential. Specifically, the eigenvalues of Eq. (8) are calculated by using the second-order perturbation theory. Then, in step (vi), $n_{well}(B)$ and $n_{bulk}(B)$ are calculated by using Eqs. (14) and (16), respectively. In step (vii), $V_0(B)$ is calculated by using Eq. (18). In steps (viii) and (ix), solving the gap equation (10), we get $\Gamma_0(B,T)$ and $2n_{\max}^{(2)}(B,T)$. In step (x), we check whether the thus-obtained $2n_{\max}^{(2)}(B,T)$ is consistent with the trial one that is given in step (iv). The steps (iv) – (x) are repeated until the consistency between the trial and thus-obtained values of $2n_{\max}^{(2)}(B,T)$ is achieved. After getting the consistent value of $2n_{\max}^{(2)}(B,T)$, we check whether Eq. (11) holds or not by using values of $2n_{\max}^{(2)}(B,T)$ and $\Gamma_0(B,T)$. The steps (ii) – (xi) are repeated until Eq. (11) holds. Thus, we can obtain the values of $\lambda$, $N_s(B,T)$ and $\Gamma_0(B,T)$ for each $B$ and $T$.

**III. RESULTS AND DISCUSSION**

Figure 4 shows the magnetic field and temperature dependences of the superconducting gap. The superconducting gap drops very abruptly from a finite value to zero as $T$ or $B$ approaches to the transition temperature or critical magnetic field. This is due to the fact that there is no solution that satisfies Eqs. (10), (11) and (20) simultaneously over the transition temperature or over the critical magnetic field. This means that no superconducting state exists over the transition temperature or over the critical magnetic field. Thus, the reproduction of the first-order phase transition observed experimentally [47,48] is reproduced by the present scheme. Figure 5 shows the magnetic field dependence of the superconducting gap with experimental results [48]. The calculated results well capture the characteristics of the experimental ones.

A key point in describing the first-order phase transition is that the magnetic field dependence of the attractive interaction is taken into consideration. Namely, if the mechanism of how the attractive interaction changes with a magnetic field is not considered, then the superconducting gap decreases smoothly and approaches to zero with increasing $T$ or $B$ [39]. On the other hand, the magnetic field dependence of the attractive interaction is taken into consideration, so that the first-order phase transition is successfully reproduced as shown in Figs. 4 and 5.

Next, let us discuss the magnetic field dependence of $V_0(B)$. Figure 6 shows the magnetic field dependence of $V_0(B)$. It is found from Fig. 6 that $V_0(B)$ gradually decreases with increasing $B$, which causes the decrease in $\Gamma_0(B,T)$ as shown in Figs. 4 and 5. Since there is no $\Gamma_0(B,T)$, $N_s(B,T)$ and $\lambda$ that satisfies Eqs. (10), (11) and (20) simultaneously as mentioned above, there are no calculated results for the attractive interaction in the high magnetic fields above the critical magnetic field. This will be discussed later. Figure 7 shows the magnetic field dependence of $N_s(B,T)/n_{\max}^{(2)}(B,T)$. This ratio represents the fraction of superconducting electrons near the surface to the total superconducting electrons. In other words, the ratio $N_s(B,T)/n_{\max}^{(2)}(B,T)$ indicates the fraction of superconducting electrons that are subject to the attraction interaction weaken by the screening effect ($V_0(B)_{well}$). The ratio $N_s(B,T)/n_{\max}^{(2)}(B,T)$ increases with $B$ as shown in Fig. 7. This means that although the total number of superconducting electrons $2n_{\max}^{(2)}(B,T)$ in the sample decreases with $B$ as shown in Fig. 8, superconducting electrons tend to gather near the surface. Due to the increase in $N_s(B,T)/n_{\max}^{(2)}(B,T)$, $V_0(B)$ decreases with $B$ as shown in Fig. 6. It should be noted that that the ratio $N_s(B,T)/n_{\max}^{(2)}(B,T)$ also represents the fraction of superconducting electrons that contribute to the antimagnetic current of the Meissner effect. It is found from Fig. 7 that the fraction of superconducting electrons contributing to the diamagnetic current increases with $B$.

Figure 9 shows the magnetic field dependences of $\lambda$ and $N_s(B,T)$. In the low- and medium-magnetic-field regions, $\lambda$ decreases with increasing $B$. Since $N_s(B,T)$ is inversely proportional to $\lambda$ as shown in Eq. (20), $N_s(B,T)$ increases with $B$. Therefoe, the fraction of superconducting electrons that are subjected to the attractive interaction weaken by the screening effect increases with $B$ as shown in Fig. 7. Thus, the decrease in $\lambda$ is consistent with the increase in $N_s(B,T)/n_{max}^{(2)}(B,T)$ and with the decrease in $V_0(B)$. Conversely, the increase in $\lambda$ leads to the decrease in $N_s(B,T)$, which may cause the increase in $V_0(B)$ because the number of superconducting electrons that are subjected to weak attractive interactions decreases. However, even though $\lambda$ increases with increasing $B$ in high magnetic field region at low temperatures as shown in Fig. 9, $V_0(B)$ decrease with increasing $B$ as shown in Fig. 6. This can be understood as follows. The depth and width of the surface potential well increases with $\lambda$, which leads to the increase in the electron density near the surface. Due to the larger screening effect, $V_0(B)$ eventually decreases with $B$ as shown in Fig. 6. Thus, the increase in $\lambda$ has opposing effects on $V_0(B)$.

If $B$ further increased over the critical magnetic field, $\lambda$ would increase. As $\lambda$ increases, the number of superconducting electrons that are subjected to weak attractive interactions decreases, i.e., $N_s(B,T)$ decreases. This, in turn, prevents the reduction of the attractive interaction. Therefore, even if $\lambda$ is increased so as to reduce the attractive interaction, the above-mentioned opposite effects do not result in a reduction of $V_0(B)$ at the high magnetic field above the critical magnetic field. As a result, $V_0(B)$ cannot be reduced at the high magnetic field above the critical magnetic field, so that the solution of the simultaneous equations cannot be found out, i.e., the superconducting state does not exist. As a result, the superconducting gap disappears abruptly at the critical magnetic field as shown in Figs. 4 and 5.

## IV. CONCLUSION

We have proposed a model for the magnetic field dependence of the attractive interaction between superconducting electrons. The proposed model states that the surface potential well produced by the penetrating magnetic field leads to the high electronic density region, which causes the magnetic field dependence of the attractive interaction. The depth of the surface potential well increases with increasing the magnetic field, so that the attractive interaction decreases due to the screening effect caused by the increase in the electronic density. We have also developed a calculation scheme to solve the gap equation of the sc-CDFT with considering the proposed model for the magnetic field dependence of the attractive interaction. Specifically, the gap equation of the sc-CDFT is solved in conjunction with both the equation for the energy balance and formula for the London penetration depth. It is shown that the simultaneous solution of these three equations does not exist at high magnetic fields above the critical magnetic field due to introducing the magnetic field dependence of the attractive interaction. Thus, it is found that the first-order phase transition observed in aluminum can be reproduced by using the sc-CDFT with the proposed model for the magnetic field dependence of the attractive interaction.

**ACKNOWLEDGEMENTS**

This work was partially supported by Grant-in-Aid for Scientific Research (No. 18K03510 and No. 18K03461) of the Japan Society for the Promotion of Science.

**REFERENCES**

[1] L. N. Oliveira, E. K. U. Gross and W. Kohn, Phys. Rev. Lett **60**, 2430 (1988).

[2] M. Lüders, M. A. L. Marques, N. N. Lathiotakis, A. Floris, G. Profeta, L. Fast, A. Continenza, S. Massidda, and E. K. U. Gross, Phys. Rev. B **72**, 024545 (2005).

[3] M. A. L. Marques, M. Lüders, N. N. Lathiotakis, Gianni Profeta, A. Floris, L. Fast, A. Continenza, E. K. U. Gross, and S. Massidda, Phys. Rev. B **72**, 024546 (2005).

[4] T. Kreibich and E. K. U. Gross, Phys. Rev. Lett. **86**, 2984 (2001).

[5] A. Floris, G. Profeta, N. N. Lathiotakis, M. Lüders, M. A. L. Marques, C. Franchini, E. K. U. Gross, A. Continenza, and S. Massidda, Phys. Rev. Lett. **94**, 037004 (2005).

[6] G. Profeta, C. Franchini, N. N. Lathiotakis, A. Floris, A. Sanna, M. A. L. Marques, M. Lüders, S. Massidda, E. K. U. Gross, and A. Continenza, Phys. Rev. Lett. **96**, 047003 (2006).

[7] A. Sanna, C. Franchini, A. Floris, G. Profeta, N. N. Lathiotakis, M. Lüders, M. A. L. Marques, E. K. U. Gross, A. Continenza, and S. Massidda, Phys. Rev. B **73**, 144512 (2006).

[8] A. Floris, A. Sanna, S. Massidda, and E. K. U. Gross, Phys. Rev. B **75**, 054508 (2007).

[9] A. Sanna, G. Profeta, A. Floris, A. Marini, E. K. U. Gross, and S. Massidda, Phys. Rev. B **75**, 020511(R) (2007).

[10] J. Quintanilla, K. Capelle, and L. N. Oliveira, Phys. Rev. B **78**, 205426 (2008).

[11] P. Cudazzo, G. Profeta, A. Sanna, A. Floris, A. Continenza, S. Massidda, and E. K. U. Gross, Phys. Rev. B **81**, 134506 (2010).

[12] C. Bersier, A. Floris, A. Sanna, G. Profeta, A. Continenza, E. K. U. Gross, and S. Massidda, Phys. Rev. B **79**, 104503 (2009).

[13] G. Stefanucci, E. Perfetto, and M. Cini, Phys. Rev. B **81**, 115446 (2010).

[14] R. Akashi, K. Nakamura, R. Arita, and M. Imada, Phys. Rev. B **86**, 054513 (2012).

[15] R. Akashi and R. Arita, Phys. Rev. B **88**, 014514 (2013).

[16] F. Essenberger, A. Sanna, A. Linscheid, F. Tandetzky, Gianni Profeta, P. Cudazzo, and E. K. U. Gross, Phys. Rev. B **90**, 214504 (2014).

[17] G. Csire, B. Újfalussy, J. Cserti, and B. Győrffy, Phys. Rev. B **91**, 165142 (2015).

[18] F. Essenberger, A. Sanna, P. Buczek, A. Ernst, L. Sandratskii, and E. K. U. Gross, Phys. Rev. B **94**, 014503 (2016).

[19] J. A. Flores-Livas, A. Sanna, and E. K. U. Gross, Eur. Phys. J. B **89**, 63 (2016).

[20] M. Monni, F. Bernardini, A. Sanna, G. Profeta, and S. Massidda, Phys. Rev. B **95**, 064516 (2017).

[21] J. A. Flores-Livas, A. Sanna, A. P. Drozdov, L. Boeri, G. Profeta, M. Eremets, and S.

Goedecker, Phys. Rev. Mater. **1**, 024802 (2017).

[22] K. Higuchi, E. Miki and M. Higuchi, J. Phys. Soc. Jpn. **86**, 064704 (2017).

[23] A. Sanna, A. Davydov, J. K. Dewhurst, S. Sharma and J. A. Flores-Livas, Eur. Phys. J. B **91**, 177 (2018).

[24] M. Lüders, P. Cudazzo, G. Profeta, A. Continenza, S. Massidda, A. Sanna and E. K. U. Gross, J. Phys.: Condens. Matter **31**, 334001 (2019).

[25] G. Marini, P. Barone, A. Sanna, C. Tresca, L. Benfatto, and G. Profeta, Phys. Rev. Mater. **3**, 114803 (2019).

[26] C. Pellegrini, H. Glawe, and A. Sanna, Phys. Rev. Mater. **3**, 064804 (2019).

[27] K. Higuchi and M. Higuchi, JPS Conf. **30**, 011066 (2020).

[28] M. Kawamura, Y. Hizume, and T. Ozaki, Phys. Rev. B **101**, 134511 (2020).

[29] T. Nomoto, M. Kawamura, T. Koretsune, R. Arita, T. Machida, T. Hanaguri, M. Kriener, Y. Taguchi, and Y. Tokura, Phys. Rev. B **101**, 014505 (2020).

[30] A. Davydov, A. Sanna, C. Pellegrini, J. K. Dewhurst, S. Sharma, and E. K. U. Gross, Phys. Rev. B **102**, 214508 (2020).

[31] K. Higuchi and M. Higuchi, J. Phys. Commun. **5**, 095003 (2021).

[32] W. Kohn, E. K. U. Gross and L. N. Oliveira, J. de Phyique (Paris) **50**, 2601 (1989).

[33] K. Higuchi, K. Koide, T. Imanishi and M. Higuchi, Int. J. Quantum Chem. **113**, 709 (2013).

[34] A. Linscheid, A. Sanna, A. Floris, and E. K. U. Gross, Phys. Rev. Lett. **115**, 097002 (2015).

[35] A. Linscheid, A. Sanna, F. Essenberger, and E. K. U. Gross, Phys. Rev. B **92**, 024505 (2015).

[36] A. Linscheid, A. Sanna, and E. K. U. Gross, Phys. Rev. B **92**, 024506 (2015).

[37] K. Higuchi, H. Niwa, and M. Higuchi, J. Phys. Soc. Jpn. **86**, 104705 (2017).

[38] K. Higuchi, N. Matsumoto, Y. Kamijo and M. Higuchi, Phys. Rev. B **102**, 014515 (2020).

[39] K. Higuchi, N. Matsumoto, Y. Kamijo and M. Higuchi, J. Phys.: Condens. Matter 33, 435602 (2021).

[40] M. Higuchi and K. Higuchi, Phys. Rev. B **69**, 035113 (2004).

[41] K. Higuchi and M. Higuchi, Phys. Rev. A **79**, 022113 (2009).

[42] M. Higuchi and K. Higuchi, Phys. Rev. A. **81**, 042505 (2010).

[43] K. Higuchi and M. Higuchi, Phys. Rev. B. **82**, 155135 (2010).

[44] M. Higuchi and K. Higuchi, Computational and Theoretical Chemistry **1003**, 91 (2013).

[45] P. G. De Gennes, *Superconductivity of Metals and Alloys* (WA Benjamin. Inc., New York, 1966).

[46] F. London and H. London, Proc. Roy. Soc. (London) A **149** 71 (1935).

[47] D. H. Douglass Jr, Phys. Rev. Lett. **7**, 14 (1961).

[48] R. Meservey and D. H. Douglass Jr, Phys. Rev. **135** A24 (1964).

[49] M. P. Marder, *Condensed Matter Physics* (Wiley, New York, 2000) Ch. 6.
[50] S. Raimes, *Many-Electron Theory* (North-Holland pub., 1972, London).

Figure captions

Figure 1: (a) Schematic diagram of the system under consideration. (b) Profiles of the surface potential wells with $k_y = \pm k_F$ for several cases of $B$ ($B$ = 1, 20, 40, 60, 7and 80 G). The value of $\lambda$ is fixed at 50 (nm) in calculating these profiles.

Figure 2: Schematic diagram of the model for the magnetic field dependence of the attractive interaction.

Figure 3: Calculation procedure.

Figure 4: Magnetic field and temperature dependences of the superconducting gap.

Figure 5: Magnetic field dependence of the superconducting gap with experimental results [49].

Figure 6: Magnetic field dependence of the propose model for the magnetic field dependence of the attractive interaction $V_0(B)$.

Figure 7: Magnetic field dependence of $N_s(B,T)/n_{max}^{(2)}(B,T)$. The ratio $N_s(B,T)/n_{max}^{(2)}(B,T)$ indicates the fraction of superconducting electrons that are subject to the attraction interaction weaken by the screening effect.

Figure 8: Magnetic field dependence of the total number of superconducting electrons $2n_{max}^{(2)}(B,T)$.

Figure 9: Magnetic field dependences of $\lambda$ and $N_s(B,T)$.

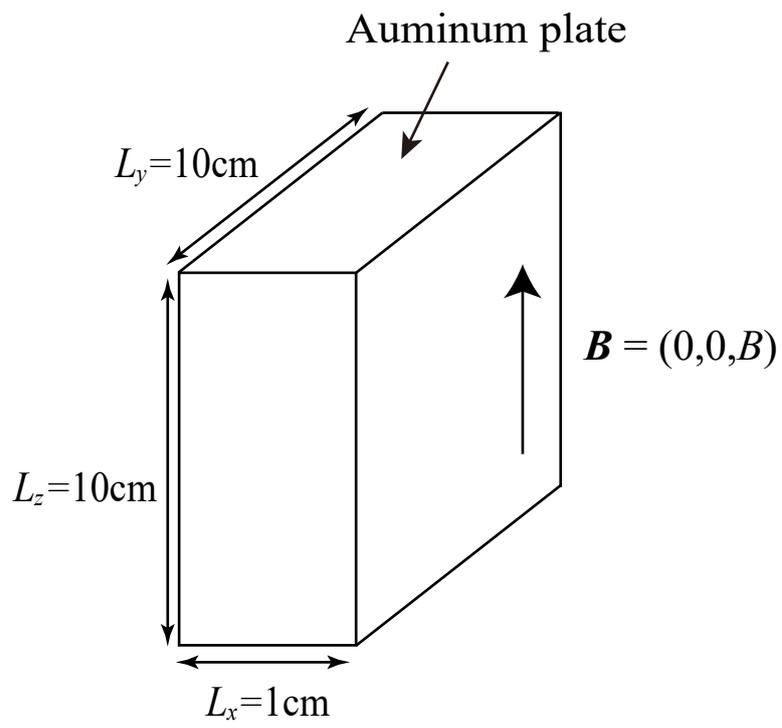

Fig. 1(a)

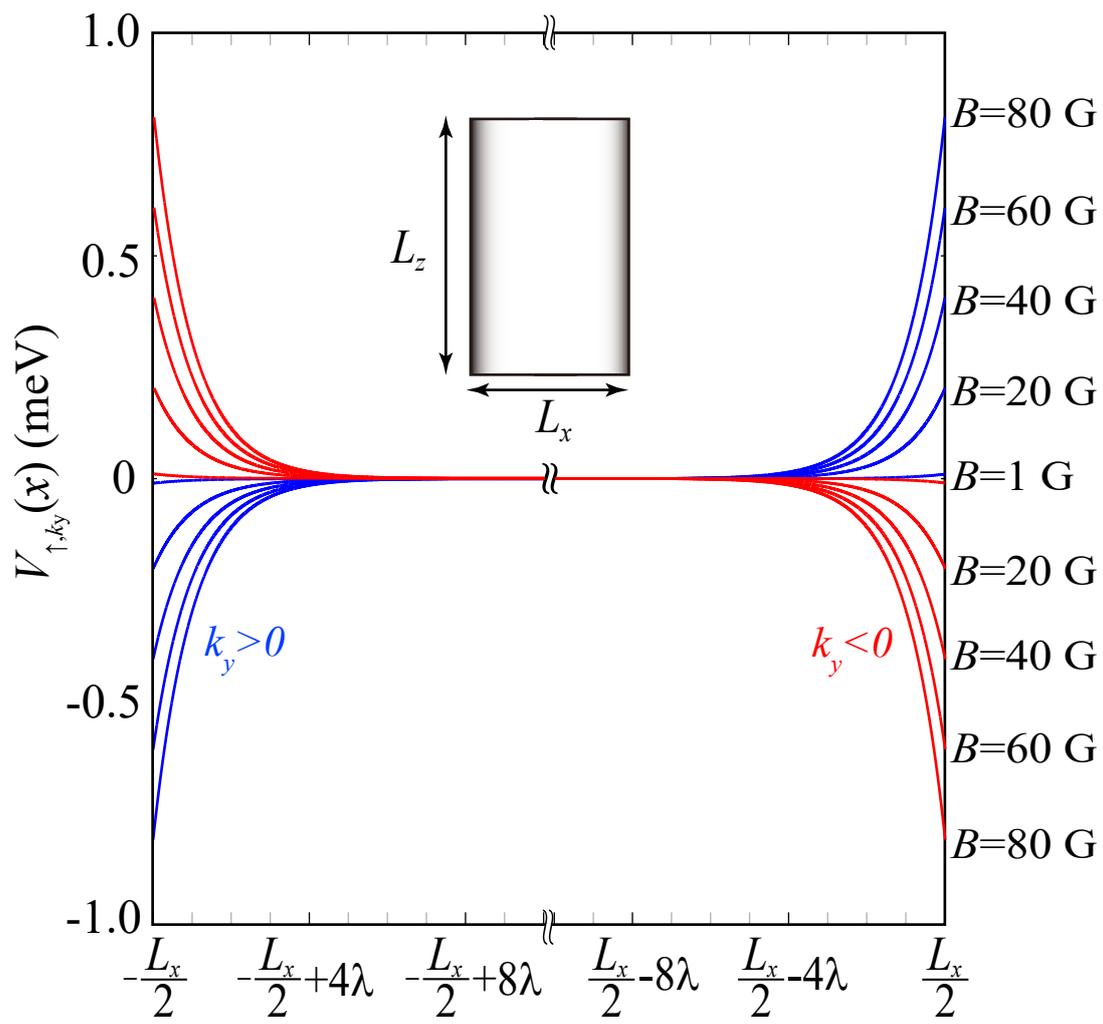

Fig. 1(b)

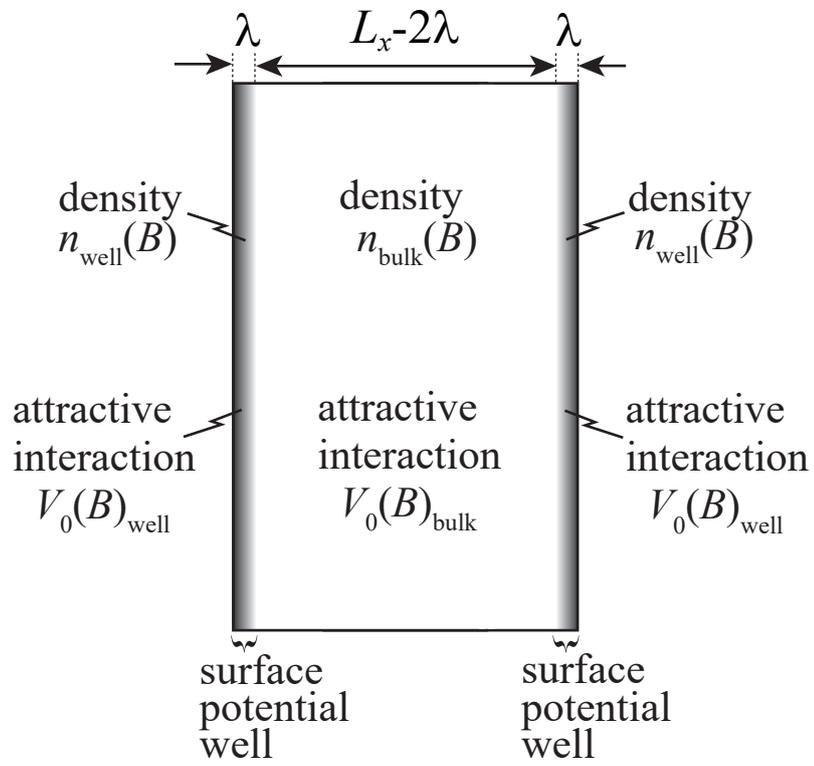

Fig. 2

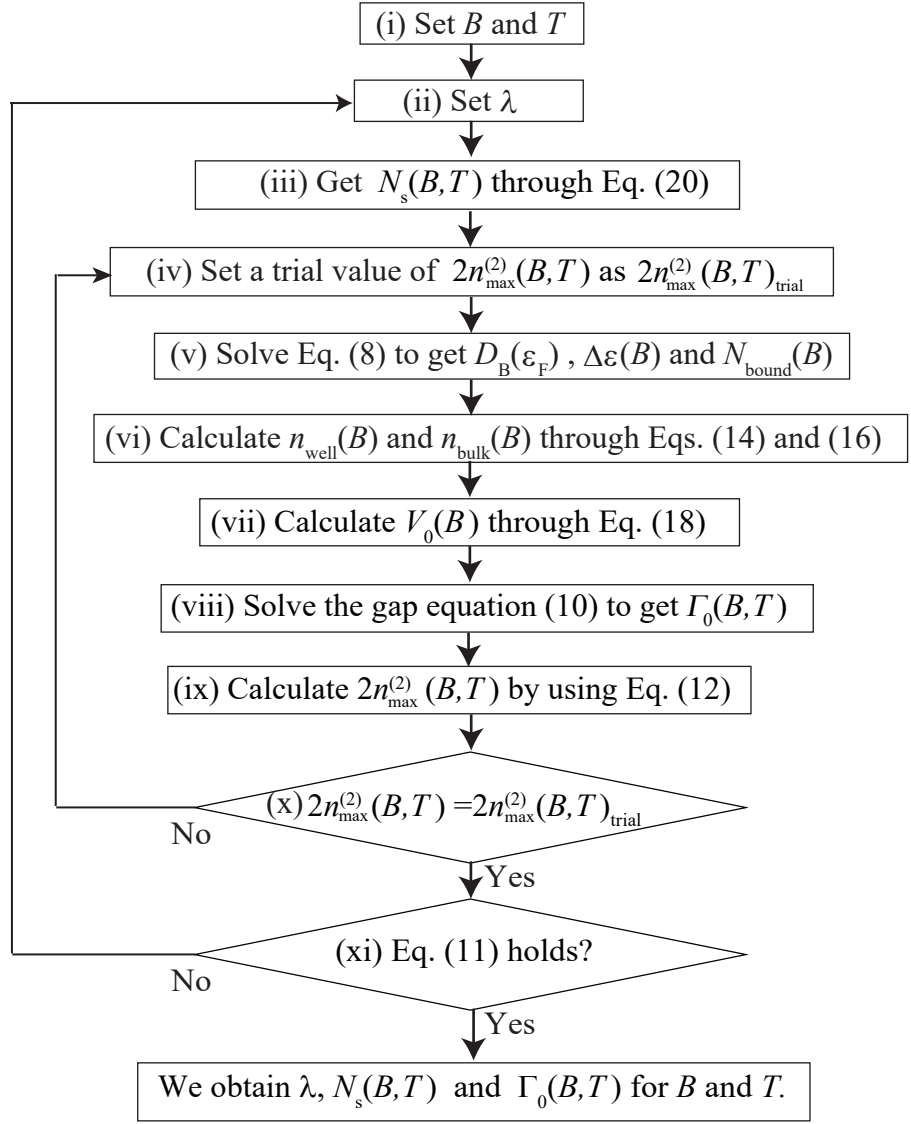

Fig. 3

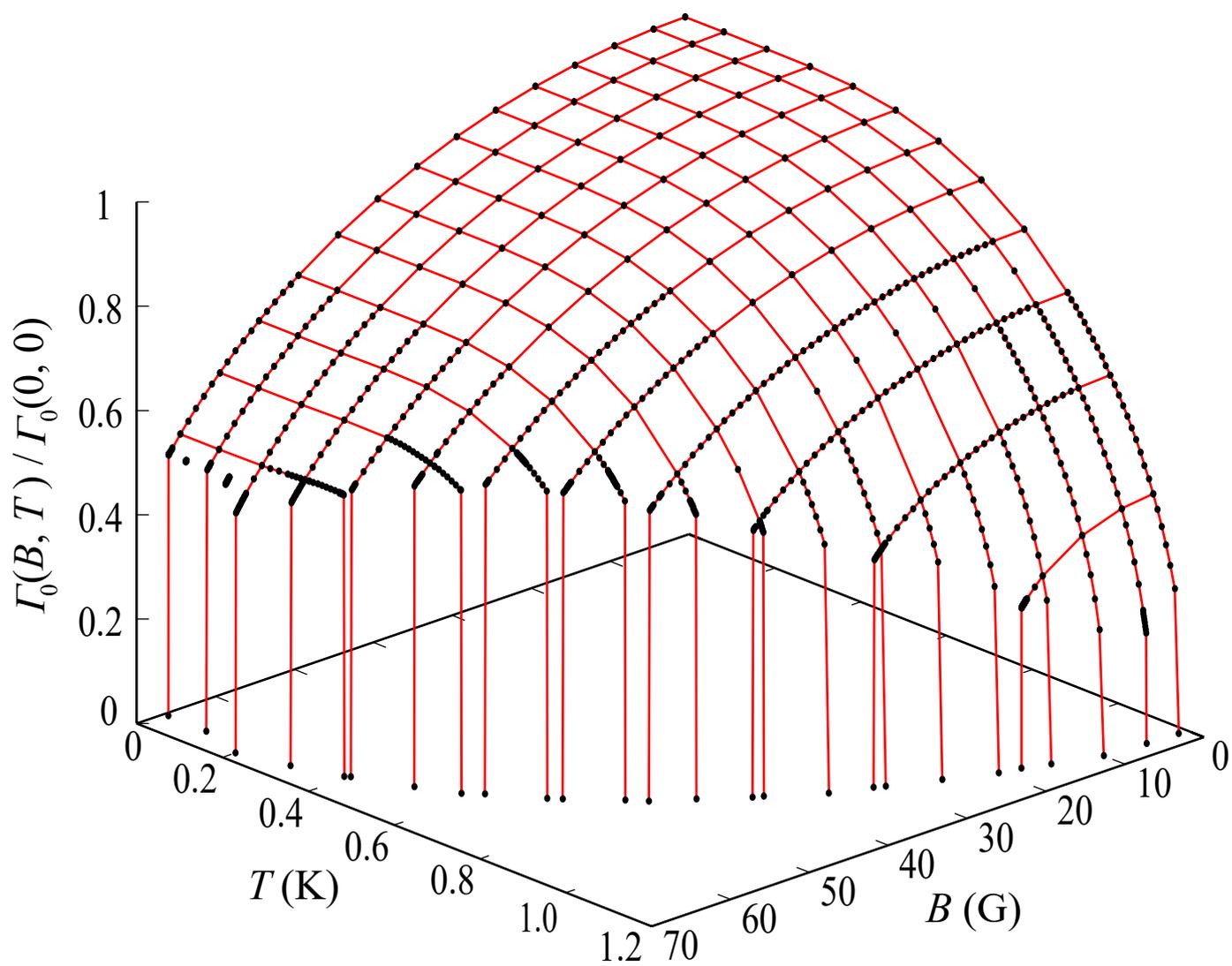

Fig. 4

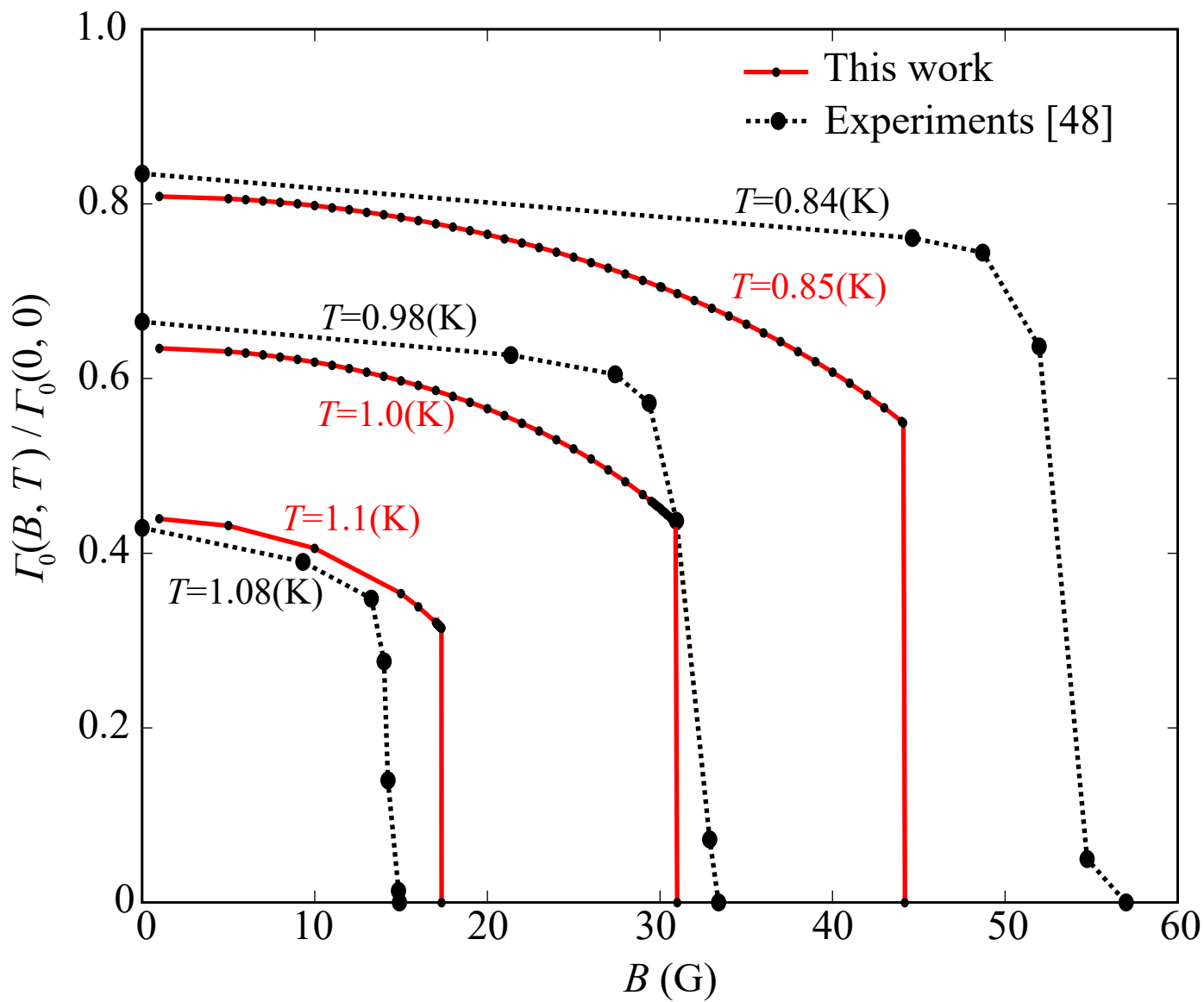

Fig. 5

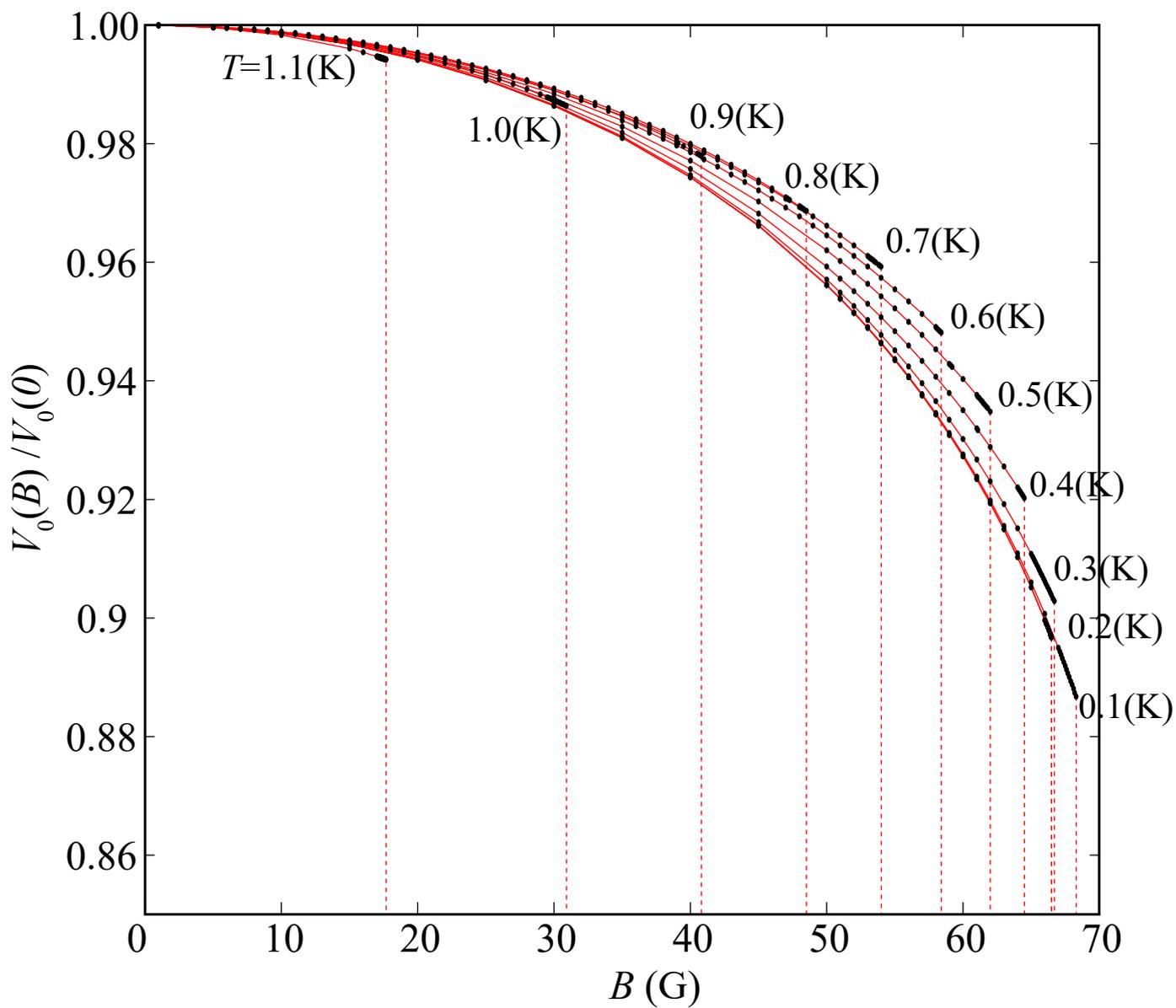

Fig. 6

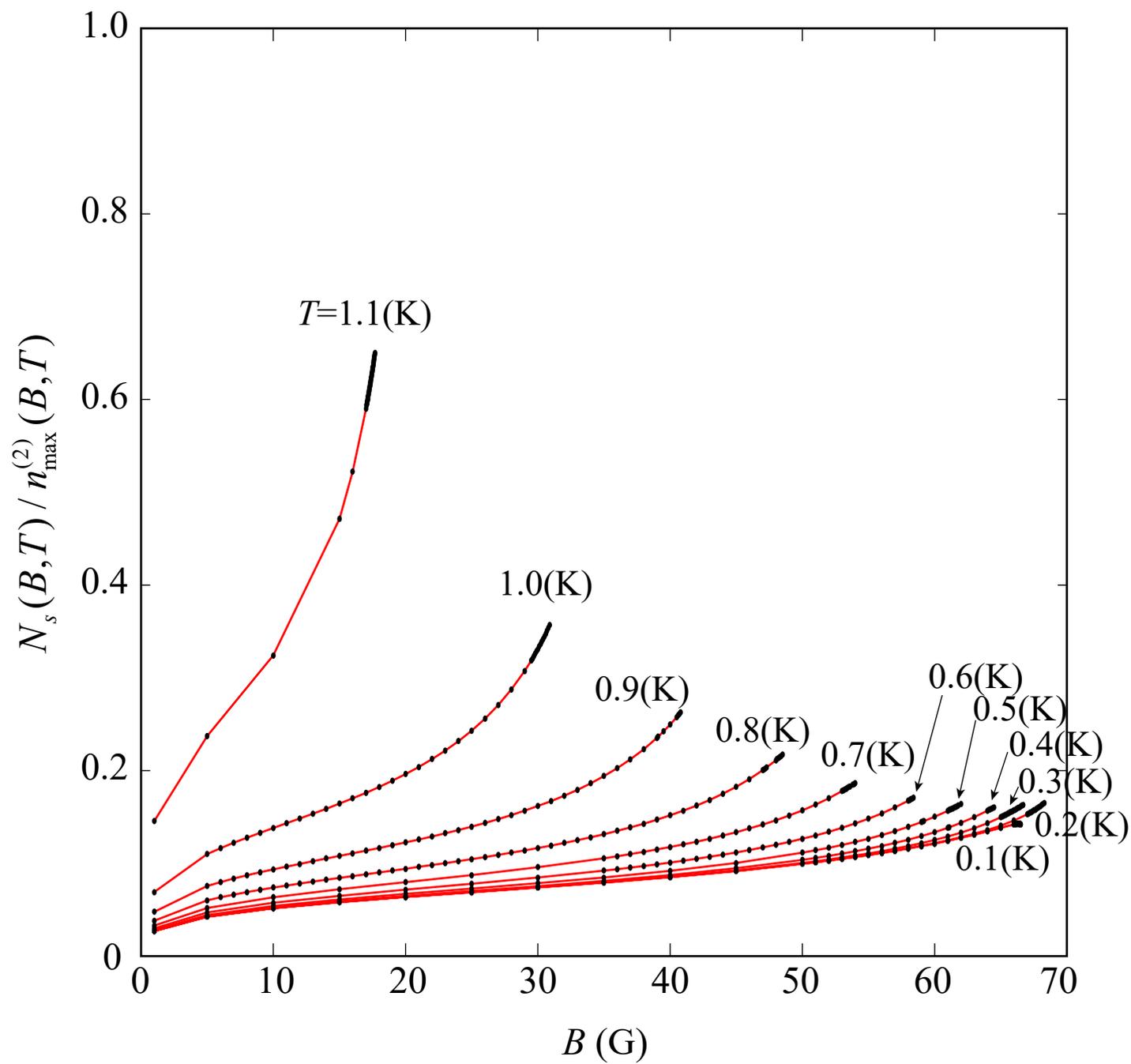

Fig. 7

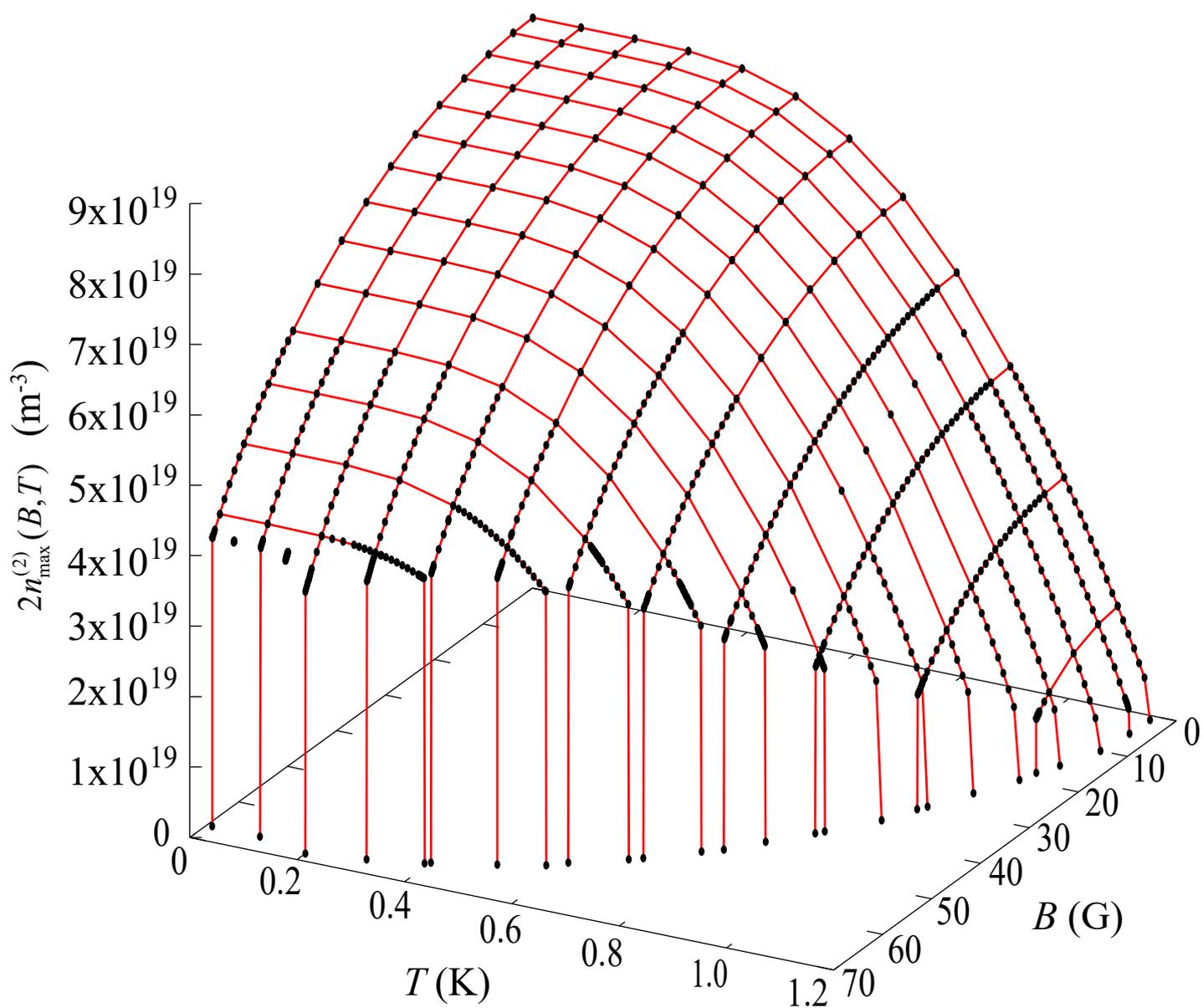

Fig. 8

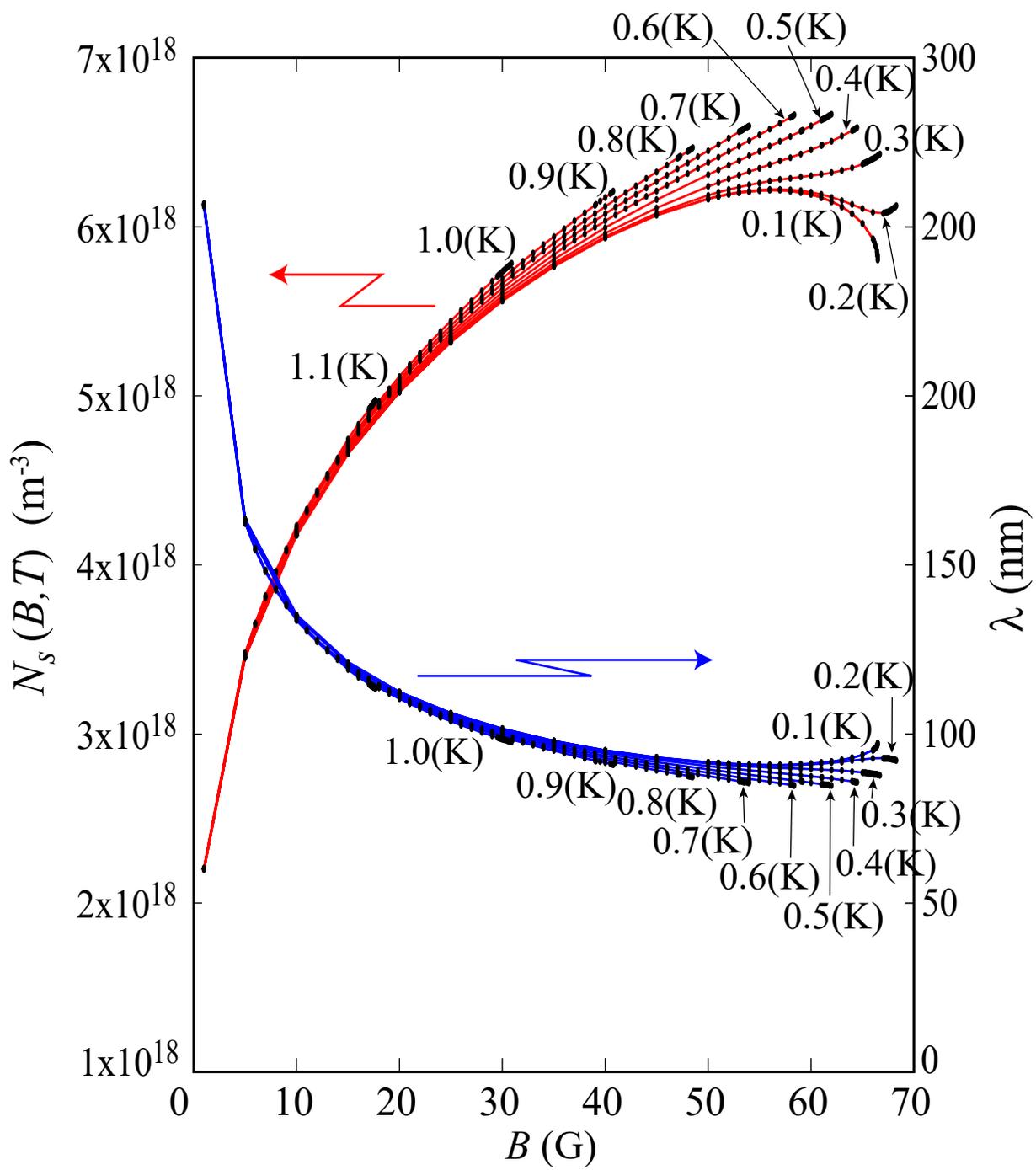

Fig. 9